\documentclass{article}
\usepackage[margin=1in]{geometry}
\usepackage{amsmath, amssymb, graphics, setspace}
\usepackage{tikz}
\usepackage{float}
\usetikzlibrary{positioning}
\usepackage{hyperref}
\usepackage[style=ieee, citestyle=numeric-comp, backend=biber, url=false]{biblatex} 
\usepackage{multirow}
\usepackage{caption}
\usepackage{subcaption}
\usepackage{graphicx}

\newcommand{\mathsym}[1]{{}}
\newcommand{\unicode}[1]{{}}
\usepackage{boldline}

\title{Reheating study of Mexican-Hat-type Potentials}
\author{Sudhava Yadav$^1$, Akash Yadav$^1$ and K.K. Venkataratnam$^{1,\footnotemark{}}$ 
\\$^1$ Department of Physics, Malaviya National Institute of Technology Jaipur, Jaipur 302017, India}
\begin{document}
\maketitle
\begin{abstract}
{We study reheating in Mexican-Hat–type potentials, emphasizing the role of the post-inflationary equation-of-state parameter($\overline{\omega }_{\text{re}}$) in shaping observable predictions. By exploring the allowed range of $\overline{\omega }_{\text{re}}$, we derive reheating temperature, e-fold counts, and inflationary observables, showing that the conventional Mexican-Hat model satisfies Planck18+BK18+BAO constraints on $n_s$ and r. The analysis underscores reheating as a critical link between theoretical potentials and CMB data. In addition, the holographic Mexican-Hat realization is examined as a benchmark, with our results mapping its phenomenological boundaries. This work illustrates how reheating studies sharpen constraints and guide refinements of unified inflationary scenarios.}
\end{abstract}
\footnotetext{*Corresponding author}
\section{Introduction}\label{S1}
The famous and exciting theory of inflation
\cite{guth_inflationary_1981,starobinsky_new_1980,linde_new_1982,linde_chaotic_1983,riotto_inflation_2002} was originally proposed to offer persuasive solutions to known problems of homogeneity and flatness of Big Bang. This theory also offers a logical justification for the genesis of the near scale-invariant primordial perturbation spectrum. During inflation, these quantum perturbations are inflated to cosmological scales and they subsequently re-enter the Hubble radius to originate the cosmic structures, as seen in Cosmic Microwave Background(CMB) anisotropies\cite{mukhanov_quantum_1981,starobinsky_dynamics_1982,guth_quantum_1985}. \\
The universe is left in a cold and frozen state at the end of inflationary epoch. Following this expansion phase, the universe must transit to a thermalised state at a very high temperature for standard cosmological evolution. This transit is termed as reheating \cite{turner_coherent_1983,traschen_particle_1990,albrecht_reheating_1982,kofman_reheating_1994,kofman_towards_1997,drewes_kinematics_2013,allahverdi_reheating_2010}. This phase of reheating is parameterized by temperature\(\left(T_{\text{re}}\right)\), duration\(\left(N_{\text{re}}\right.\)) and effective EoS($\overline{\omega }_{\text{re}}$). These parameters can 
be related to inflationary ones and are indirectly bounded by recent CMB data\cite{martin_observing_2015,martin_first_2010,dai_reheating_2014,martin_inflation_2006,adshead_inflation_2011,mielczarek_reheating_2011,cook_reheating_2015,yadav2024modified,goswami_reconciling_2018,yadav2024mutated}.
The reheating dynamics is sensitively dependent on the shape of the inflationary potential\cite{dai_reheating_2014} and its interactions with other fields, which makes the potential form a subject of central importance.\\
A lot of potentials are studied in the recent past\cite{martin_encyclopaedia_2013}.
A rich and well-motivated framework is provided in this context by scalar potentials exhibiting spontaneous symmetry breaking\cite{goldstone1961field}, especially, the Mexican-Hat potentials.
The Mexican-Hat form, most famously embodied in the double-well inflation (DWI) model, has long served as a cornerstone in early-universe cosmology due to its capacity to capture symmetry-breaking dynamics and remain compatible with modern observational bounds\cite{MartinSchwarz2003,Destri2007,Bostan2018}. These potentials occur spontaneously in many field-theoretic models, its characteristic feature involves a local maximum at origin with a ring of degenerated minima at finite values of field. Such potentials can provide different predictions for observed variables in cosmology by supporting inflationary scenarios in which the field evolves inside small-field regimes or close to the hilltop (origin) or small-field region, resulting in different predictions for observed variables. Furthermore, topological defects like domain walls can arise as a result of the spontaneous breakdown of discrete symmetries present in these potentials\cite{goldstone1961field, peter1995surface,linde1994topological,ringeval2000equation, witten1985superconducting,peter1993spontaneous,carter1994supersonic, peter2000fermionic,ringeval2001fermionic,vilenkin1994topological}, which have significant ramifications for the early universe's history. The details of the potential's structure and the kinetics of symmetry breaking are closely related to the existence and stability of these topological defects.\\
 More recently, this evocative potential has gained renewed traction in unified frameworks that aim to link the epochs of primordial inflation and late-time cosmic acceleration. For instance, Jiménez-Aguilar’s holographic spacetime-foam model naturally gives rise to a Mexican-Hat potential as a consequence of demanding that a scalar field—representing the classical energy of spacetime foam within the Hubble volume—exactly satisfies the cosmological field equations; this provides a unified treatment of both early- and late-time acceleration without invoking separate inflaton and dark-energy fields\cite{jimenez2023unified}.\\
Our current study delves deep into the reheating dynamics driven by these Mexican-Hat-type potentials, highlighting theoretical consistency and their alignment with recent observations, Planck18+BK18+BAO data\cite{aghanim_planck_2020,akrami_planck_2020-1,tristram2022improved}. Enforcing the condition \(T_{\text{re}}> 100\) GeV on account of generation of weak-scale dark matter and letting the EoS parameter to cover the range ($-\frac{1}{3} \le \overline{\omega }_{\text{re}} \le 1$), we made use of the relation between reheating and inflationary parameters along with observational viable values of scalar spectral index(\(n_{\text{s}}\)), amplitude of scalar power spectrum(\(A_{\text{s}}\)) and tensor-to-scalar ratio(\(r\)) to constrain the models.\\
The layout of this article is arranged as follows: Section~\ref{S1} introduces the motivation and background of the study. 
Section~\ref{S2} contains a concise discussion on the reheating parameterization. Section~\ref{S3} contains reheating dynamics of various Mexican-Hat-type potentials along with the main results and analysis, followed by concluding remarks in the final section.\\
Throughout this work, we adopt natural units with $\hbar = c = 1$. For reference, the reduced Planck mass is taken as $M_{P} = \sqrt{\tfrac{1}{8 \pi G}} \simeq 2.435 \times 10^{18} \, \text{GeV}$. We also use the matter–radiation equality redshift $z_{\text{eq}} \approx 3400$, the effective number of relativistic degrees of freedom $g_{\text{re}} \approx 100$ \cite{dai_reheating_2014}, and the present Hubble parameter $H_{0} = 100h \, \text{km}\, s^{-1}\, \text{Mpc}^{-1}$ with $h = 0.68$\cite{aghanim_planck_2020,akrami_planck_2020-1}.\\

\begin{section}{Reheating Formalism}\label{S2}
We commence with a concise overview of the reheating formalism, formulated within the framework adopted in Ref.\cite{goswami_reconciling_2018,yadav2024modified,yadav2024mutated}, in order to establish the notation and assumptions employed in the present analysis. The single-field inflation models with inflaton field($\phi$) having potential $V(\phi)$ slowly arise with parameters:
\begin{equation} \label{1}
H^2=\frac{1}{3M_P^2}V(\phi),
\end{equation}
\begin{equation} \label{2}
\epsilon =\frac{M_P^2}{2}\left(\frac{V'(\phi)}{V(\phi)}\right)^2,
\end{equation}
\begin{equation} \label{3}
\eta =M_P^2\left(\frac{V''(\phi)}{V(\phi)}\right),
\end{equation}
where \((')\) signifies derivative w.r.t \(\phi\). Further, we have expressions for inflation parameters:

\begin{equation}\label{4}
\text{   }n_s=1-6\epsilon +2\eta.
\end{equation}

\begin{equation}\label{5}
r=16\epsilon.
\end{equation}

Between the end of inflation(subscript end) and the point at which a mode k crosses the Hubble horizon(subscript k), the expression for e-foldings is
\begin{equation}\label{6}
\Delta N_k=\frac{1}{M_P^2}\int_{\phi _{\text{end}}}^{\phi _k} \frac{V(\phi)}{V'(\phi)} \, d\phi.
\end{equation}
Now, the expressions for reheating parameters in terms of inflationary parameters are
\begin{equation} \label{7}
\ln  (T_{\text{re}})=\frac{3~\left(1+\overline{\omega }_{\text{re}}\right)}{3~ \overline{\omega }_{\text{re}}-1}\left\{\ln \left(\rho _{\text{eq}}^{\frac{1}{4}}\right)-\ln
\left(1+z_{\text{eq}}\right)-\ln  \frac{k}{a_o}-\Delta N_k+\ln  H_k\right\}
-\frac{1}{3~\overline{\omega }_{\text{re}}-1}\ln \left(\frac{3}{2}V_{end}\right)-\frac{1}{4}\ln \left(\frac{\pi ^2}{30}g_{re}\right),
\end{equation}

\begin{equation}\label{8}
N_{\text{re}}=\frac{1}{3~\overline{\omega }_{\text{re}}-1}\ln \left(\frac{3}{2}V_{\text{end}}\right)+\frac{4}{3~\overline{\omega }_{\text{re}}-1}\left\{\ln
\left(\frac{k}{a_o}\right)+\Delta N_k+\ln \left(1+z_{\text{eq}}\right)-\ln \left(\rho _{\text{eq}}^{\frac{1}{4}}\right) -\ln  H_k \right\}.
\end{equation}

\end{section}
\section{Mexican-Hat-type potentials}\label{S3}
\subsection{Double well inflation(DWI)}\label{S3.1}
\begin{figure}[H]
    \centering
    \includegraphics[width=0.5\linewidth]{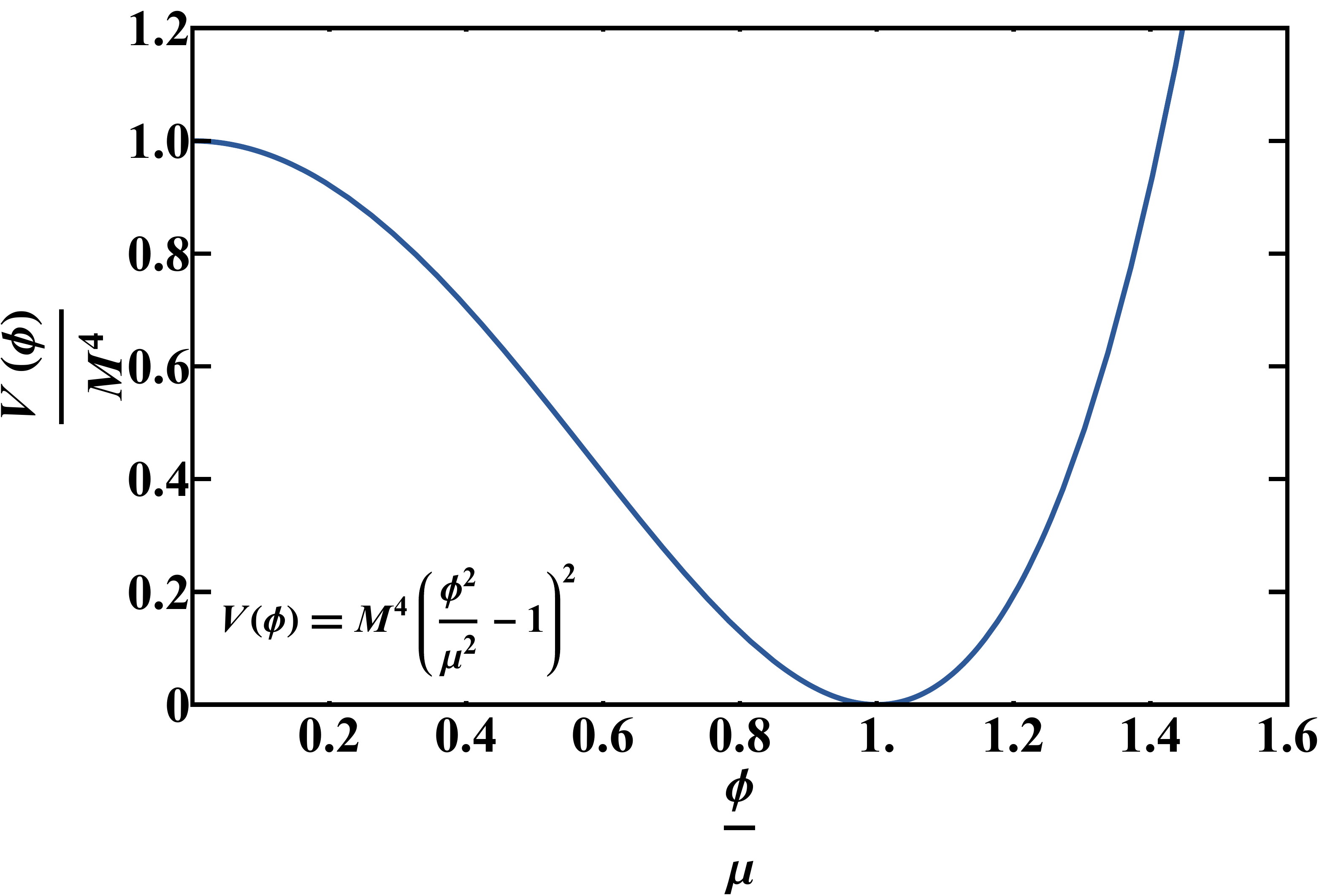}
    \caption{Variation of DWI with $\frac{\phi}{\mu}$}
    \label{F1}
\end{figure}

\begin{figure}[p]
\centering
{\includegraphics[width=\textwidth]{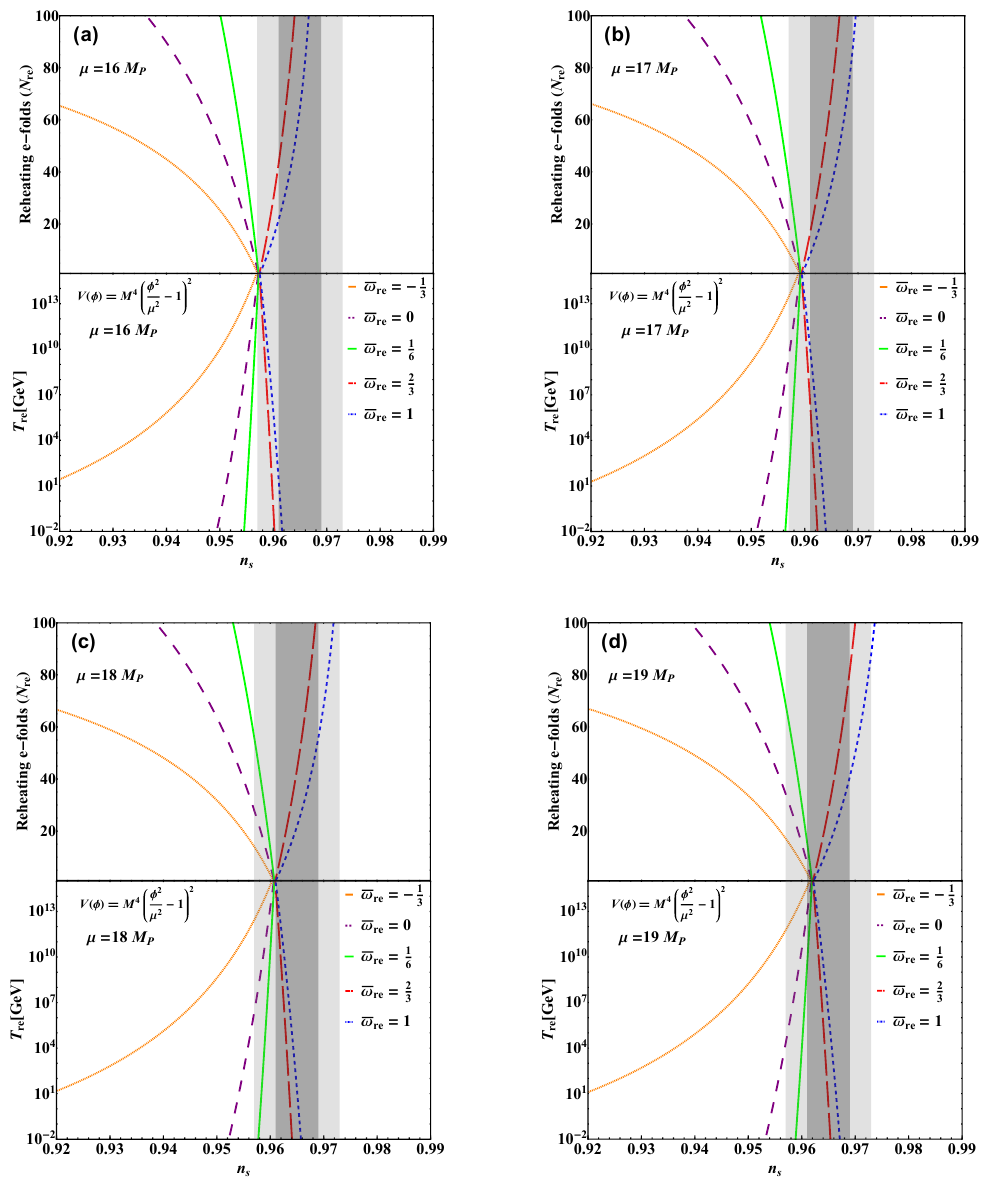}}
\caption{The plot panels showing variation of \(T_{\text{re}}\) and \(N_{\text{re}}\) w.r.t \(n_{s }\) for double well inflation for multiple values of EoS parameter : \(\overline{\omega }_{\text{re}}\)=\(-\frac{1}{3}\)(orange, dotted), 0(purple, medium dashed), \(\frac{1}{6}\)(green, solid), \(\frac{2}{3}\)(red, large dashed), 1(blue, small dashed). The gray bands represent the \(\text{1$\sigma $}\) and \(\text{2$\sigma $}\) ranges of \(n_s\) from Planck18(TE,EE,TT+Low E+Lensing)\cite{aghanim_planck_2020,akrami_planck_2020-1}.}
    \label{F2}
\end{figure}

\begin{figure}[H]
      \centering
 \includegraphics[width=0.5\linewidth]{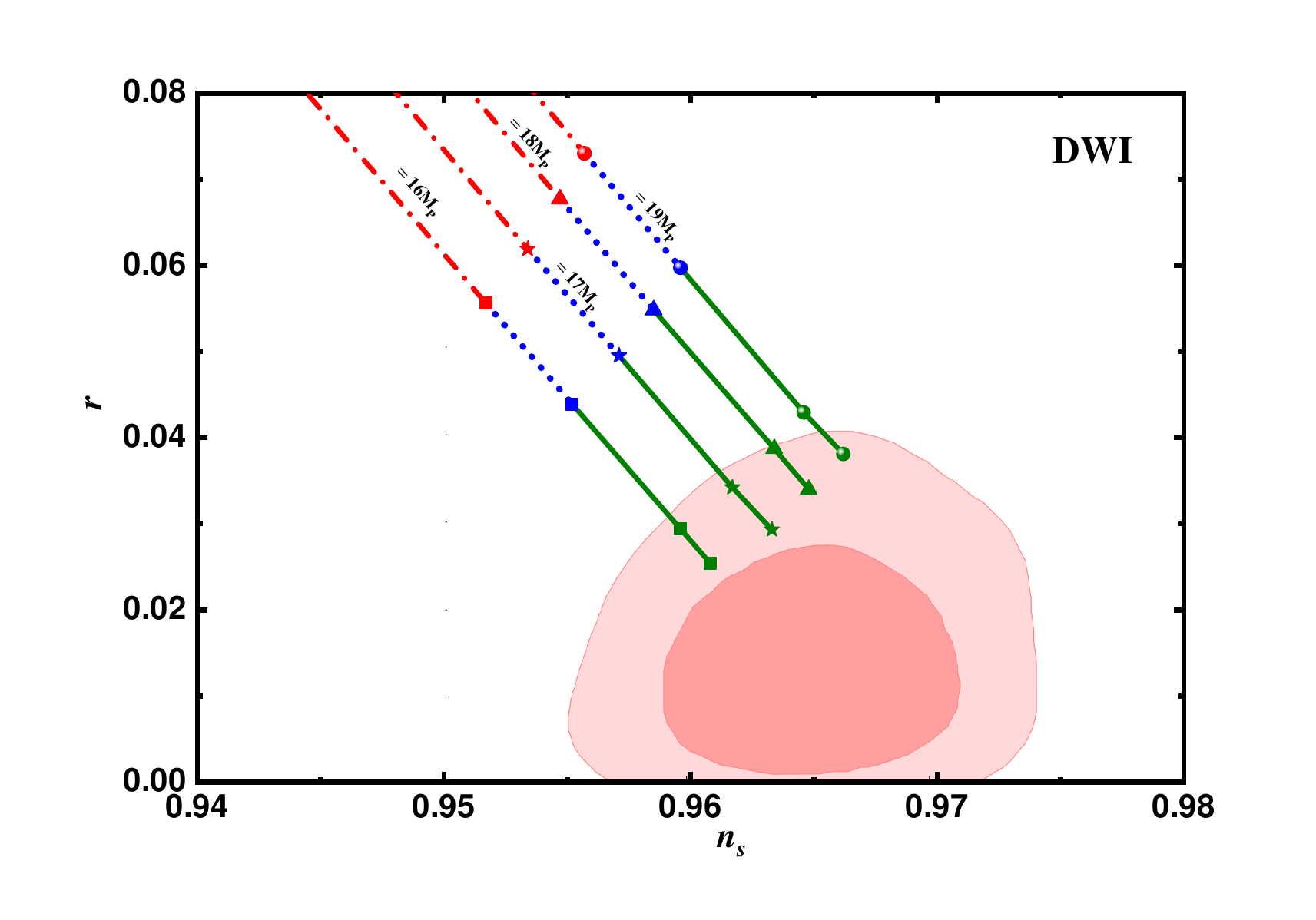}
      \caption{The (\(n_s\),r) predictions for different mass scale for DWI with different {$\mu$} and \(\overline{\omega }_{\text{re} }\): $-\frac{1}{3} \le \overline{\omega }_{\text{re}} < 0$( red, dotted-dash), $0 \le \overline{\omega }_{\text{re}} \le \frac{1}{6}$(blue dotted), $\frac{1}{6} < \overline{\omega }_{\text{re}} \le 1$(green, solid ). The peach outlines in the backdrop are from Planck18+BK18+BAO observations \cite{tristram2022improved}.}
      \label{F3}
  \end{figure}

The double well inflation potential has the form
\begin{equation} \label{9}
V(\phi ) = M^4[\left(\frac{\phi}{\mu}\right)^2 -1]^2,
\end{equation}
where $\mu$ represents the mass scale and M is the normalization term. The DWI potential has been extensively studied across cosmology and field theory—for topological defect formation, serving as a model for dynamical symmetry breaking, and as a basis for topological inflation caused by spontaneous breaking of discrete symmetry\cite{goldstone1961field, peter1995surface, linde1994topological,ringeval2000equation,witten1985superconducting, peter1993spontaneous,carter1994supersonic,peter2000fermionic,ringeval2001fermionic,vilenkin1994topological}. We will be studying here the reheating dynamics driven by this potential and highlighting its theoretical alignment with Planck18+BK18+BAO observations\cite{aghanim_planck_2020,akrami_planck_2020-1,tristram2022improved}. The variation of DWI with $\frac{\phi}{\mu}$ is shown in Figure \ref{F1}. Since the potential is symmetric under $\phi \rightarrow -\phi$, only the $\phi > 0$ area is shown. 
The slow-roll inflationary parameters for the DWI potential can be stated as\\
\begin{equation} \label{10}
\text{$\epsilon $ =}\frac{8M_P^2 \phi^2 }{(\mu^2 -\phi ^2)^2},
\end{equation}
\begin{equation} \label{11}
\text{$\eta $ = }-\frac{4M_P^2 (\mu^2 -3\phi ^2)}{(\mu^2 -\phi ^2)^2},
\end{equation}
\begin{equation}\label{12}
n_s=1-\frac{8M_P^2 (\mu^2 +3\phi ^2)}{(\mu^2 -\phi ^2)^2},
\end{equation}
 \begin{equation}\label{13}
r =\frac{128M_P^2 \phi^2 }{(\mu^2 -\phi ^2)^2}.
\end{equation}
Now, imposing the condition \(\epsilon =1\), which defines the end of inflation, to obtain $\phi_{\text{end}}$ and reviving the pivot scale as used by the Planck collaboration mission, \(\frac{k_*}{a_o}=0.05 Mpc^{-1}\), consider the same mode \(k_*\), which will cross the hubble radius at a certain point  where the field gets the value \(\phi _*\) during inflation. The remaining e-folds after this crossing are
\begin{equation} \label{14}
\Delta N_*\simeq \frac{1}{4M_P^2}( \frac{\phi ^2}{2} - u^2 ln(\phi))\\
\end{equation}
Moreover, for DWI model 
\begin{equation} \label{15}
H_*=8\pi  M_P\sqrt{\frac{A_sM_P^2 \phi^2 }{(\mu^2 -\phi ^2)^2}}.
\end{equation}
\begin{equation} \label{16}
V_{\text{end}}(\phi )=M^4[\left(\frac{\phi_{\text{end}}}{\mu}\right)^2 -1]^2=\frac{3H_*^2M_P^2(\left(\frac{\phi_{\text{end}}}{\mu}\right)^2 -1)^2}{(\left(\frac{\phi}{\mu}\right)^2 -1)^2}
\end{equation}
Now, inserting these derived inflationary parameters for DWI in the expressions \ref{7} and \ref{8} to obtain reheating parameters. We have utilized  \(\text{
 }A_{s }=2.1\times 10^{-9}\) and  \(\rho
_{\text{eq}}^{\frac{1}{4}} = 10^{-9}\)GeV \cite{aghanim_planck_2020,akrami_planck_2020-1} for our computation. The parameters \(T_{\text{re}}\) and \(N_{\text{re}}\) are mapped against \(n_s\) for DWI in Figure \ref{F2}, exploring four representative values of the mass scale $\mu$ and a spectrum of EoS parameter($\overline{\omega }_{\text{re}}$) with Planck18 \(2\sigma\) and \(1\sigma\) bounds on \(n_s\) in background. The curves illustrate that the $\overline{\omega }_{\text{re}}$ branches begin to drift beyond acceptable $n_s$ limits as $\mu$ decreases. For smaller $\mu$, the allowed $N_{\mathrm{re}}$ values are generally larger, implying longer reheating durations for a given $n_s$, while larger $\mu$ shifts the viable $T_{\mathrm{re}}$ range toward higher temperatures. This behavior underscores the sensitivity of reheating predictions to the inflationary energy scale and post-inflationary expansion history. The intersection of all curves marks instantaneous reheating: minimal \(N_{\text{re}}\), maximal \(T_{\text{re}}\), and it is independent of $\overline{\omega }_{\text{re}}$.\\
On enforcing the condition \(T_{\text{re}} \geq 100\) GeV, the admissible bounds on various inflationary parameters are obtained and compiled in Table \ref{T1}. The table highlights how increasing $\mu$ modifies the inflationary predictions. The obtained numerical bounds on $n_s$ and r for DWI from Table \ref{T1} are visually represented in Figure \ref{F3} with Planck18+BK18+BAO contour bound\cite{tristram2022improved} in background. Negative $\overline{\omega }_{\text{re}}$ (dotted-dash red) tends to push predictions toward lower $n_s$ and higher $r$, corresponding to longer reheating phases, whereas positive $\bar{\omega}_{\mathrm{rh}}$ (solid green) compresses the predictions toward higher $n_s$ and smaller $r$, characteristic of a stiffer reheating equation of state.
The Planck18+BK18+BAO contours already rule out a substantial region of parameter space. On systematically varying the EoS parameter, we found that the data contours already rule out substantial regions of parameter space, helping us find that the values of $\mu$ lie in the observationally viable window for DWI. This model for chosen values of $\mu$ lies in observational range for higher $\omega _{\text{re}}$.  Particularly, for $\mu \sim 17M_P$ the model covers the viable window for most extensive range of $\overline{\omega }_{\text{re}}$.\\

\begin{table}[!t]
\caption{The admissible bounds on inflationary parameters for DWI by demanding $T_{re} \geq 100GeV$ for different mass scale} \label{T1}
    \centering
    \begin{tabular}{|c|c|c|c|c|c|}
        \hline
         \multirow{5}{*}{$\mu$ = $16 M_P$} & Effective EoS & $n_s$ & $\Delta N_*$ & $r$ \\
        \cline{2-5}
        & $-\frac{1}{3} \le \overline{\omega }_{\text{re}} \le 0$ & $0.923 \le n_s \le 0.952$ & $26.08 \le \Delta  N_* \le 46.9$ & $0.151  \ge r \ge 0.055$ \\
         \cline{2-5}
        & $0 \le \overline{\omega }_{\text{re}} \le \frac{1}{6}$ & $0.952 \le n_s \le 0.955$ & $46.9 \le \Delta  N_* \le 52.78$ & $0.055  \ge r \ge 0.044$ \\
         \cline{2-5}
        & $\frac{1}{6} \le \overline{\omega }_{\text{re}} \le \frac{2}{3}$ & $0.955 \le n_s \le 0.959$ & $52.78 \le \Delta  N_* \le 63.29$ & $0.044  \ge r \ge 0.030$ \\
        \cline{2-5}
        & $\frac{2}{3} \le \overline{\omega }_{\text{re}} \le 1$ & $0.959 \le n_s \le 0.961$ & $63.29 \le \Delta  N_* \le 67.35$ & $0.030  \ge r \ge 0.025$ \\
        \hlineB{4}
        \multirow{4}{*}{$\mu$ = $17 M_P$}
        & $-\frac{1}{3} \le \overline{\omega }_{\text{re}} \le 0$ & $0.924 \le n_s \le 0.953$ & $26.11 \le \Delta  N_* \le 46.94$ & $0.159  \ge r \ge 0.062$ \\
         \cline{2-5}
        & $0 \le \overline{\omega }_{\text{re}} \le \frac{1}{6}$ & $0.953 \le n_s \le 0.957$ & $46.94 \le \Delta  N_* \le 52.82$ & $0.062  \ge r \ge 0.049$ \\
         \cline{2-5}
        & $\frac{1}{6} \le \overline{\omega }_{\text{re}} \le \frac{2}{3}$ & $0.957 \le n_s \le 0.962$ & $52.82 \le \Delta  N_* \le 63.35$ & $0.049  \ge r \ge 0.034$ \\
        \cline{2-5}
        & $\frac{2}{3} \le \overline{\omega }_{\text{re}} \le 1$ & $0.962 \le n_s \le 0.963$ & $63.35 \le \Delta  N_* \le 67.42$ & $0.034  \ge r \ge 0.029$ \\
        \hlineB{4}
         \multirow{4}{*}{$\mu$ = $18 M_P$} 
        & $-\frac{1}{3} \le \overline{\omega }_{\text{re}} \le 0$ & $0.925 \le n_s \le 0.955$ & $26.12 \le \Delta  N_* \le 46.98$ & $0.167  \ge r \ge 0.067$ \\
         \cline{2-5}
        & $0 \le \overline{\omega }_{\text{re}} \le \frac{1}{6}$ & $0.955 \le n_s \le 0.958$ & $46.98 \le \Delta  N_* \le 52.86$ & $0.067  \ge r \ge 0.054$ \\
         \cline{2-5}
        & $\frac{1}{6} \le \overline{\omega }_{\text{re}} \le \frac{2}{3}$ & $0.958 \le n_s \le 0.963$ & $52.86 \le \Delta  N_* \le 63.40$ & $0.054  \ge r \ge 0.038$ \\
        \cline{2-5}
        & $\frac{2}{3} \le \overline{\omega }_{\text{re}} \le 1$ & $0.963 \le n_s \le 0.964$ & $63.40 \le \Delta  N_* \le 67.48$ & $0.038  \ge r \ge 0.034$ \\
         \hlineB{4}
         \multirow{4}{*}{$\mu$ = $19 M_P$} 
        & $-\frac{1}{3} \le \overline{\omega }_{\text{re}} \le 0$ & $0.925 \le n_s \le 0.955$ & $26.14 \le \Delta  N_* \le 47.00$ & $0.175  \ge r \ge 0.073$ \\
         \cline{2-5}
        & $0 \le \overline{\omega }_{\text{re}} \le \frac{1}{6}$ & $0.955 \le n_s \le 0.959$ & $47.00 \le \Delta  N_* \le 52.90$ & $0.073  \ge r \ge 0.059$ \\
         \cline{2-5}
        & $\frac{1}{6} \le \overline{\omega }_{\text{re}} \le \frac{2}{3}$ & $0.959 \le n_s \le 0.964$ & $52.90 \le \Delta  N_* \le 63.44$ & $0.059  \ge r \ge 0.043$ \\
        \cline{2-5}
        & $\frac{2}{3} \le \overline{\omega }_{\text{re}} \le 1$ & $0.964 \le n_s \le 0.966$ & $63.44 \le \Delta  N_* \le 67.53$ & $0.043  \ge r \ge 0.038$ \\
        \hline
    \end{tabular}
\end{table}

\subsection{Holographic spacetime foam unified inflation(HFUM)}\label{S3.2}
\begin{figure}[!h]
\centering
\includegraphics[width=0.5\textwidth]{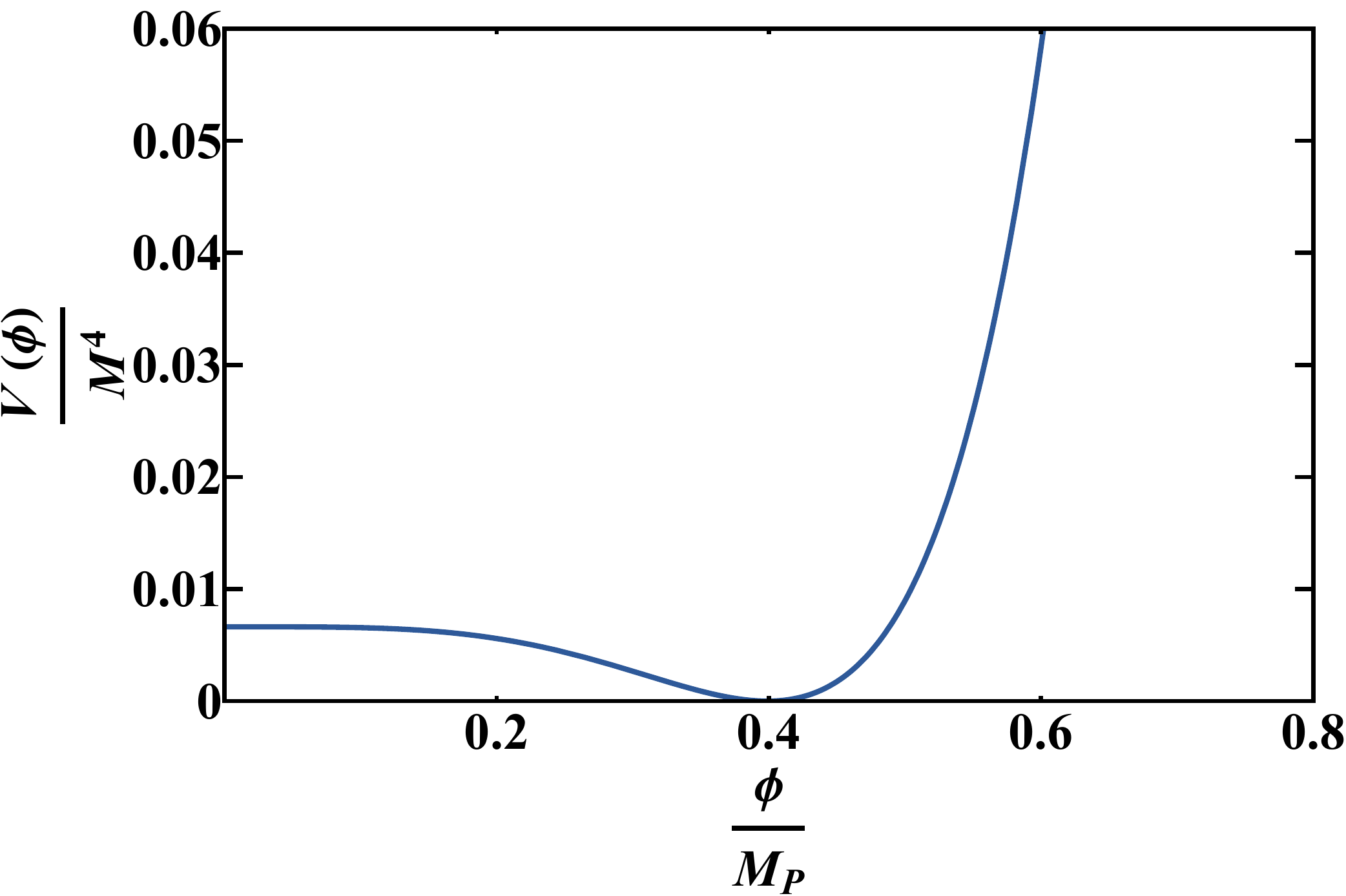}
\caption{The potential versus $\frac{\phi}{M_P}$ plot for HFUM.}
\label{F4} 
\end{figure}

 The Holographic spacetime foam unified inflation is a well-motivated model. It has a distinctive topography, possibly with steeper slopes or asymmetric minima, unlike the standard Mexican-Hat potential. It has the form
\cite{jimenez2023unified}
\begin{equation} \label{28}
V(\phi )=M^4 \left(\frac{ \pi ^2}{3} \left(\frac{\phi ^6}{M_P^6}-\frac{3 \phi ^4}{4 M_P^4 \pi }\right)+\frac{1}{48 \pi }\right),
\end{equation}
where $(M)^4=m^2M_P^2$. We are considering a specific realization of the holographically reconstructed complex scalar potential, as originally proposed in Ref.\cite{jimenez2023unified}. In particular, we are focusing on the special scenario where the scaling parameter $\gamma=\frac{2}{3}$. This specific case minimizes the inflaton mass, resulting in analytic simplifications and contains the qualitative features of the inflation-dark energy unified scenario. We will be studying this unique potential in reheating context. The variation of potential in Eq.\ref{28} with $\frac{\phi}{M_P}$ is shown in Figure \ref{F4}.\\
For the HFUM potential, the slow-roll inflationary characteristics are expressed as\\
\begin{equation} \label{10}
\text{$\epsilon $ =}\frac{1152 M_P^2 \pi ^4 \phi ^6}{\left(M_P^2-2 \pi  \phi ^2\right)^2 \left(M_P^2+4 \pi  \phi ^2\right)^2},
\end{equation}
\begin{equation} \label{11}
\text{$\eta $ = }\frac{48 M_P^2 \pi ^2 \left(-3M_P^2\phi^2+10 \pi  \phi ^4\right)}{\left(M_P^2-2 \pi  \phi ^2\right)^2 \left(M_P^2+4 \pi  \phi ^2\right)},
\end{equation}
\begin{equation}\label{12}
n_s=\frac{M_P^8+4 M_P^6 (1-72 \pi ) \pi  \phi ^2-12 M_P^4 \pi ^2 (1+16 \pi ) \phi ^4-32 M_P^2 \pi ^3 (1+96 \pi ) \phi ^6+64 \pi ^4 \phi ^8}{\left(M_P^2-2
\pi  \phi ^2\right)^2 \left(M_P^2+4 \pi  \phi ^2\right)^2},
\end{equation}
 \begin{equation}\label{13}
r =\frac{18432 M_P^2 \pi ^4 \phi ^6}{\left(M_P^2-2 \pi  \phi ^2\right)^2 \left(M_P^2+4 \pi  \phi ^2\right)^2},
\end{equation}

 The remaining e-folds after hubble crossing by mode \(k_*\) are
\begin{equation} \label{14}
\Delta N_*\simeq \frac{M_P^4+ 8 \pi ^2 \phi ^4 -4 M_P^2 \pi \phi ^2  ln(\phi)}{96 M_P^2 \pi ^2\phi ^2}.
\end{equation}

\begin{figure}[!h]
\centering
{\includegraphics[width=\textwidth]{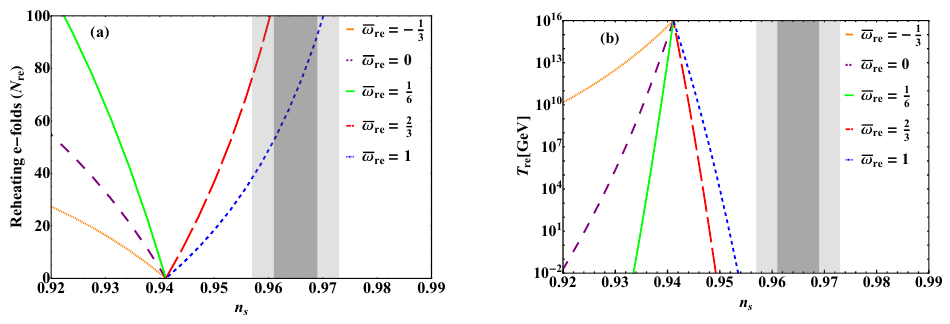}}
\caption{The plots showing variation of \(T_{\text{re}}\) and \(N_{\text{re}}\) w.r.t \(n_{s }\) for HFUM. The shaded areas and color schemes are identical to those used in Figure \ref{F2}.}
    \label{F5}
\end{figure}

\begin{figure}[!h]
\centering
{\includegraphics[width=0.6\textwidth]{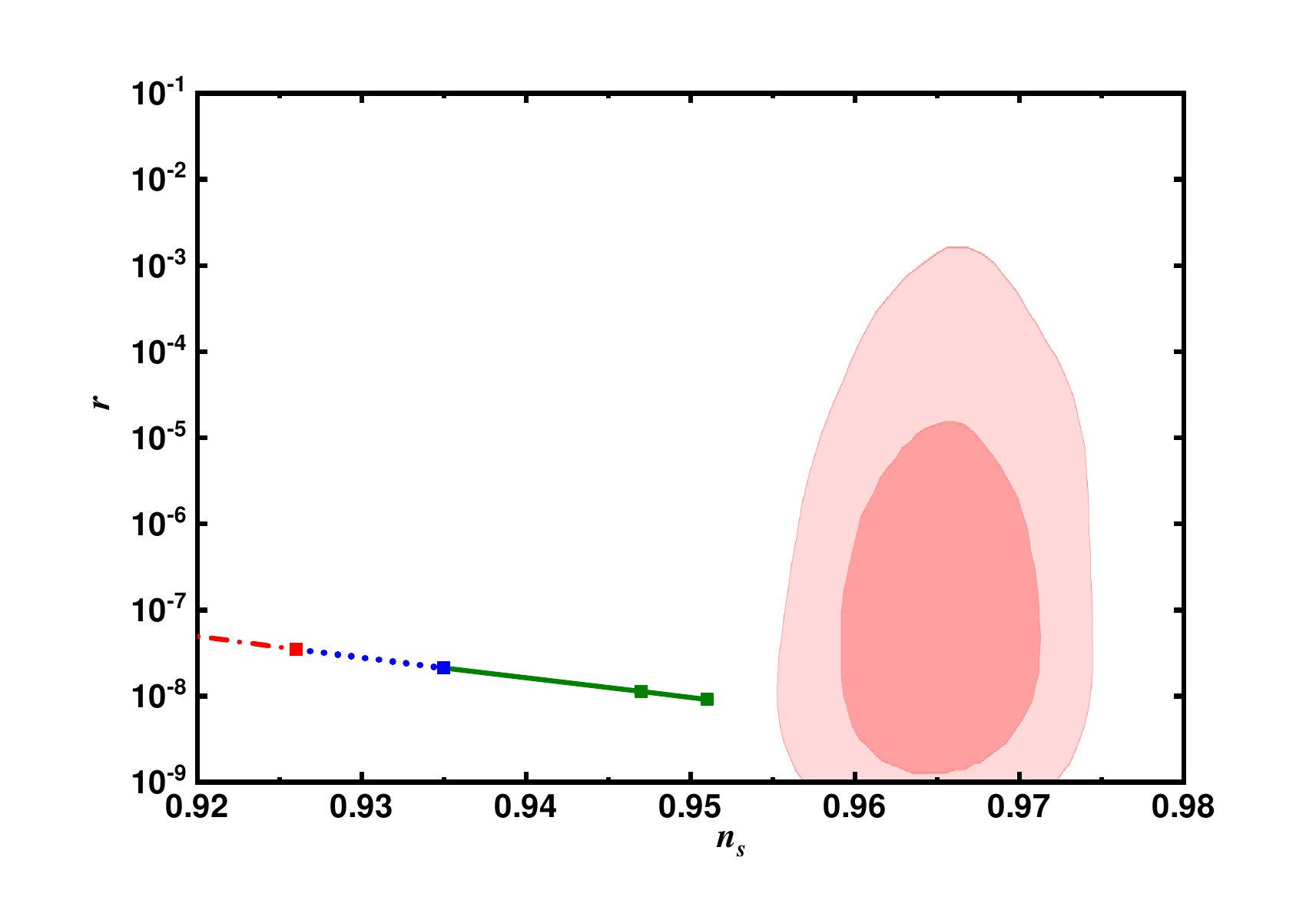}}
\caption{The (\(n_s\),r) predictions for HFUM. The contour and color schemes are identical to those used in Figure \ref{F3}.}
    \label{F6}
\end{figure}

Moreover, for HFUM model 
\begin{equation} \label{15}
H_*=\frac{96\pi^3  M_P^2\phi ^3\sqrt{A_s}}{\left(M_P^2-2 \pi  \phi ^2\right) \left(M_P^2+4 \pi  \phi ^2\right)}.
\end{equation}

\begin{equation} \label{16}
V_{\text{end}}(\phi )=M^4 \left(\frac{ \pi ^2}{3} \left(\frac{\phi_{end} ^6}{M_P^6}-\frac{3 \phi_{end} ^4}{4 M_P^4 \pi }\right)+\frac{1}{48 \pi }\right)=\frac{3H_*^2M_P^2\left(\frac{ \pi ^2}{3} \left(\frac{\phi_{end} ^6}{M_P^6}-\frac{3 \phi_{end} ^4}{4 M_P^4 \pi }\right)+\frac{1}{48 \pi }\right)}{\left(\frac{ \pi ^2}{3} \left(\frac{\phi ^6}{M_P^6}-\frac{3 \phi ^4}{4 M_P^4 \pi }\right)+\frac{1}{48 \pi }\right)}
\end{equation}
Now, inserting these derived inflationary parameters for HFUM in the expressions \ref{7} and \ref{8} to obtain reheating parameters. The parameters \(T_{\text{re}}\) and \(N_{\text{re}}\) are mapped against \(n_s\) for HFUM in Figure \ref{F5} for a spectrum of EoS parameter($\overline{\omega }_{\text{re}}$).\\ 
On enforcing the condition \(T_{\text{re}} \geq 100\)GeV, the admissible bounds on various inflationary parameters are obtained and compiled in Table \ref{T2}. The obtained numerical bounds on $n_s$ and r from Table \ref{T2} are visually represented in Figure \ref{F6}.\\
\begin{table}[!h]
\caption{The permissible range for values of inflationary parameters for HFUM by demanding $T_{re} \geq 100GeV$}\label{T2}%
    \centering
    \begin{tabular}{|c|c|c|c|}
        \hline
          Effective EoS & $n_s$ & $\Delta N_*$ & $r*10^{-8}$ \\
        \hline
         $-\frac{1}{3} \le \overline{\omega }_{\text{re}} \le 0$ & $0.84 \le n_s \le 0.926$ & $16.26 \le \Delta N_* \le 37.88$ & $32.3 \ge r \ge 3.2$ \\
         \hline
         $0 \le \overline{\omega }_{\text{re}} \le \frac{1}{6}$ & $0.926 \le n_s \le 0.935$ & $37.88 \le \Delta N_* \le 43.97$ & $3.2  \ge r \ge 2.1$ \\
         \hline
         $\frac{1}{6} \le \overline{\omega }_{\text{re}} \le \frac{2}{3}$ & $0.935 \le n_s \le 0.947$ & $43.97 \le \Delta N_* \le 54.86$ & $2.1 \ge r \ge 1.1$ \\
        \hline
         $\frac{2}{3} \le \overline{\omega }_{\text{re}} \le 1$ & $0.947 \le n_s \le 0.951$ & $54.86 \le \Delta N_* \le 59.08$ & $1.1  \ge r \ge 0.9$ \\
        \hline   
    \end{tabular}  
\end{table}
On systematically varying the EoS parameter, we found that the post-inflationary evolution for this model produces a tensor-to-scalar ratio that is well within the most recent observations. The corresponding scalar spectral index predictions, however, are just beyond the data constraints. Our analysis suggests the $n_s$–tension is not fully linked to the inflationary potential itself or to post-inflationary assumptions. The contradiction is consistent over whole EoS range, suggesting that mapping the EoS range won't be adequate to compensate for the shape-driven variations in slow-roll characteristics. This makes it clearer that the issue is intrinsic to potential’s curvature and related slow-roll dynamics, not with the reheating history.

\section{Discussion and conclusion}\label{S4}
In this study, we have performed a comparative reheating study of two Mexican-Hat-type potentials: the standard symmetry-breaking form and a modified variant inspired by the holographic spacetime foam model that unifies dark energy and inflation. We studied the reheating dynamics driven by these potentials and highlighted their theoretical alignment with Planck18+BK18+BAO observations.\\
We began our reheating study with DWI potential. This potential permits a wide range of the mass scale($\mu$) values that align with the $n_s$-r observational constraints; however, our reheating analysis narrows this to a more limited, feasible interval. Within this narrowed range, the $\mu \sim 17M_P$ case stays congruent with the data across a broad spectrum of EoS parameter. Hence, establishes it as a notably stable point inside the model's parameter space. Such stability improves the predictive power of the potential, as it remains observationally viable under diverse reheating scenarios.\\
On the contrary, the reheating study of Holographic spacetime foam unified model(HFUM) yields $n_s$ values that are substantially beyond the observationally permitted range for whole range of EoS, despite its theoretical promise in bridging the gap between rapid inflationary early universe and current accelerated universe expansion. According to our reheated study, the variation in $n_s$ remains independent of post-inflationary assumptions, suggesting that the discrepancy is not a result of late-time dynamics but rather an aspect of the holographic structure itself.\\
The stark contrasts in alignment of these two Mexican-Hat-type potentials with cosmological observations can arise from shifted vacuum alignment
and steep curvature in holographic form, unlike the standard symmetric(DWI) case. This shifted alignment in holographic form alters the inflationary slow-roll dynamics and results in deviated $n_s$ values. This deviation, caused by prolonged oscillations of inflaton field and impeding transfer of energy during reheating,
featuring the model's susceptibility to corrections inspired by quantum gravity. By testing the model through the reheating phase, we establish its phenomenological boundaries and clarify the extent to which its predictions align with current data. This provides a well-defined framework for exploring future refinements—such as alternative reheating scenarios, quantum corrections, or extensions of the holographic setup—without overstating claims, while ensuring that the model remains a useful reference point in studies of inflationary unification.

\section*{Acknowledgments}
 SY would like to acknowledge the Ministry of Education, Government of India, for providing senior research fellowship. 
\printbibliography
\end{document}